\documentclass[a4paper,fleqn,usenatbib]{mnras}

\usepackage[T1]{fontenc}
\usepackage{ae,aecompl}
\usepackage{ulem}

\usepackage{amsmath}	
\usepackage{amssymb}	
\usepackage{lscape}
\usepackage{afterpage}
\usepackage{capt-of}
\usepackage{rotating}

\usepackage{graphicx}
\usepackage{epstopdf}
\DeclareGraphicsExtensions{.eps}

\usepackage{threeparttable}
\usepackage{url}
\newcommand{\be}{\begin{equation}}
\newcommand{\ee}{\end{equation}}
\newcommand{\bea}{\begin{eqnarray}}
\newcommand{\eea}{\end{eqnarray}}
\def\deg{^\circ}
\def\arcsec{^{\prime\prime}}
\def\arcmin{^\prime}

\usepackage{color}              
\usepackage{epsfig}
\usepackage{amstext}

\title[Peripheral Diffuse Emission in Bullet cluster]{`Rings' of diffuse radio emission surrounding the Bullet cluster}

\author[Paul S., et al.]{
Surajit Paul,$^{1}$\thanks{E-mail: surajit@physics.unipune.ac.in}
Abhirup Datta,$^{2,4}$ 
Siddharth Malu,$^{2}$
Prateek Gupta,$^{1}$
Reju Sam John,$^{1,3}$ \newauthor
Sergio Colafrancesco$^{5}$
\\
$^{1}$Department of Physics, SP Pune University, Pune - 411007, India\\
$^{2}$Centre of Astronomy, Indian Institute of Technology Indore, Simrol, Khandwa Road, Indore 453552, India \\
$^{3}$Department of Physics, Pondicherry Engineering College, Puducherry - 605014, India\\
$^{4}$Center for Astrophysics and Space Astronomy, Department of Astrophysical and Planetary Science, University of Colorado, \\ 
Boulder, CO 80309, USA \\
$^{5}$School of Physics, University of the Witwatersrand, Private Bag 3, WITS-2050, Johannesburg, South Africa
}

\date{Accepted XXX. Received YYY; in original form ZZZ}

\pubyear{2017}

\begin{document}
\label{firstpage}
\pagerange{\pageref{firstpage}
\pageref{lastpage}}
\maketitle

\begin{abstract}
We present the discovery of ringlike diffuse radio emission structures in the peripheral regions of the Bullet cluster 1E 0657$-$55.8. Ring formations are spanning between 1--3 Mpc away from the center of the cluster, significantly further away from the two already reported relics. Integrated fluxes of four of the sub-regions in the inner `ring' from 4.5 to 10 GHz have also been reported. To understand the possible origin of these structures, here we present a maiden attempt of numerical modelling of a 3D and realistic `bullet' like event in a full cosmological ($\Lambda$CDM) environment with N-body plus hydrodynamics code. We report a simulated `bullet' found inside a (128 Mpc)$^3$ volume simulation with a speed of 2700 km s$^{-1}$, creating a high supersonic bow shock of Mach $M=3.5$ and a clear evidence of temporal separation of dark matter and baryons, assuring no challenge to $\Lambda$CDM cosmology from the bullet event as of now. We are also able to unveil the physics behind the formation of these observed multiple shock structures. Modelled radio emissions in our simulation support a complex combination of merger-associated processes that accelerates and re-accelerates fossil and cosmic-ray electrons. With a time evolution study and the  computed radio emissions, we have shown that the ring like formation around the bullet is originated due to the interaction of the strong merger shocks with the accretion shocks at the periphery. The multiple shock structures observed are possibly originated from multiple mergers that have taken place at different times and much before the bullet event. 
\end{abstract}




\section{Introduction}
\label{intro}

Galaxy clusters, the most massive gravitationally bound objects in the universe, grow through continuous mergers and accretion processes. Most of the matter in galaxy clusters is in the form of dark matter, and the ordinary matter or baryon is present as a plasma which constitutes the Intra-Cluster Medium (ICM), owing to the high temperature of about 10$^{7}-$10$^8$ K. Mergers and collisions between these galaxy clusters produce shocks and turbulence \citep{Sarazin_2002ASSL} that accelerate ICM plasma and amplifies magnetic field by compression \citep{Iapichino_2012MNRAS} and turbulent dynamo \citep{Subramanian_2006MNRAS} and fills the whole cluster with magnetic fields. Accelerated charged particles in the ICM gyrate in the ambient magnetic fields, give rise to central extended radio synchrotron emission (radio halos) as well as peripheral elongated diffuse emission (radio relics) along the shocks \citep{Feretti_2012,Kale_2016JApA}.

Large scale structure formation process becomes non-linear and complicated in late times of its evolution with increased rate of mergers \citep{2000Natur.406..376C,2015SSRv..188...93P}. This makes it very difficult to formulate any analytical model to understand some of the observed structures and its energetics. Bullet cluster is one of such events. It is an extreme and rare merging event with very complicated physics of structure formation \citep{Clowe_2006ApJ,Kraljic_2015JCAP}. Unlike the usual situation where, baryons remain trapped inside the dark matter halos, multi-wavelength observations of bullet cluster (1E0657-56) show an offset of gravitational potential and X-ray peak, indicating a clear separation of baryons from the dark matter (DM) \citep{2006ESASP.604..723M,2002A&A...386..816B}. According to \citet{Lee_2010ApJ,Kraljic_2015JCAP}, this even provides a challenge to the $\Lambda$CDM cosmology. Observations show a shock with Mach number more than $M=3$ ahead of the bullet, indicating extreme high-speed merger. Energetically, bullet cluster has been found to be the hottest (10s of KeV) and most X-ray and SZ bright cluster compared to other similar clusters \citep{2016PASJ...68...88K,2008A&A...491..363O,2014A&A...562A..60O}. These findings suggest that bullet is a violently merging system \citep{2012PASJ...64...12A}. The energy that release during such mergers (more than $10^{65}$ erg s$^{-1}$) dissipates through shocks that thermalize the ICM and induce large-scale turbulence as well \citep{Sarazin_2002ASSL,Paul_2011ApJ,Iapichino_2010CONF}.

Cosmological structure formation shocks are mainly known to be of two kinds \citep{2003ApJ...593..599R}, the first being merger shocks, due to direct interaction of substructures, seen in several cluster mergers by now \citep{2006Sci...314..791B,Paul_2014ASInC}, most notably in the Bullet cluster \citep{2015MNRAS.449.1486S}. The shock strength of mergers are usually reported to be $\lesssim$ 3 \citep{2003ApJ...583..695G}. The second kind, referred to as accretion shocks, occur when gas falls in towards the surroundings of the clusters \citep{Bagchi_2011ApJ}. These accretion shocks have significantly higher Mach numbers than merger shocks, at $M\gg10$ \citep{John_2018arXiv,2000ApJ...542..608M,2006MNRAS.367..113P,2008MNRAS.391.1511H}, due to the gas in the outskirts of clusters, never being heated before. It has been pointed out that the energy released during  mergers create huge pressure in the cluster core and the medium eventually starts expanding supersonically similar to a blast wave \citep{Ha_2017arXiv,Paul_2012JoPConS,Sarazin_2002ASSL} that travels radially as spheroidal wave-front towards the virial radius and moves beyond \citep{Weeren_2011MNRAS,Paul_2011ApJ,Iapichino_2017MNRAS}. So, there could a third situation arise, where, merger shock interacts with the accretion shock and re-energises the ambient particles to much higher energies and can show up in non-thermal emissions with brighter than expected. Clusters are therefore likely to be surrounded by these different kinds of shocks at distances up to a few times the virial radius \citep{2003ApJ...587..514N,2008MNRAS.391.1511H} or till the accretion zone. So, the diffuse radio emissions observed at a few Mpc away from the centre of some of the clusters may be associated with these shocks, depending on the merger dynamics and merger activity of the cluster, as evidenced by X-ray morphology, the presence and structure of radio relics and their distances from the merging clusters. In this context we should mention that \citet{2007MNRAS.375...77H} have remarked that radio emission resulting from the accretion shocks may also be detected, if confined to a small region. In order to have detectable diffuse emission from accretion shocks or the interacting shocks, a significant amount of diffuse emission, and magnetic field strengths ($\sim\mu$G), combined with sufficient shock acceleration efficiency, are needed \citep{2008MNRAS.391.1511H,2001ApJ...563..660F}. Also, depending on the stage of the merger, time elapsed, radio relics may be located $\sim$ Mpcs away from the cluster merger.

So, to understand the radio emissions from merger/accretion scenario better, a full cosmological simulation with thermal and non-thermal physics is the set-up that is required. We also need numerical models to compute the possible radio synchrotron emissions from these structures. For this purpose, we have performed a cosmological simulation with the Adaptive Mesh Refinement (AMR), grid-based hybrid (N-body plus hydro-dynamical) code Enzo v.~2.2 \citep{Bryan_2014ApJS}. And further, computed radio emissions as a post process by implementing electron spectrum from Diffusive Shock Acceleration as well as Turbulent re-acceleration models to recreate the radio map of a bullet like simulated cluster. Radio emission modelling has been done as the post processing numerical analysis by using the yt-tools \citep{Turk_2011ApJS}. 

In this paper, we present the observation of unique radio structures in bullet cluster using Australia Telescope Compact Array (ATCA) at 5.5 and 9.0 GHz. Further to understand the observed complex and unusual structures, we present our first ever simulations of a modelled bullet like event in a full cosmological environment. So, after introducing in Section~\ref{intro}, we present the observation, data analysis and reported the findings from observation in Section~\ref{radio_obs}. Simulation of a bullet like event and comparison with our observations are presented in Section~\ref{model}.  Finally, discussed the results and concluded in Section~\ref{last}.

\section{Radio Observations \& Radio Images}
\label{radio_obs}

\begin{center}
\begin{table}
\caption{Summary of the ATCA observations}\label{Obs-sum}
\label{obs_journal}
\begin{tabular}{@{}lr}
\hline\hline
Co-ordinates (J2000) RA--Dec & $06^{\rm h}58^{\rm m}30^{\rm s} -55\degr57\arcmin00\arcsec$ \\ 
Primary Beam FWHM & 8.5$\arcmin$ \\
Synthesized Beam FWHM & 35.5$\arcsec\times$13.8$\arcsec$\\ &Natural Resolution (5.5 GHz) \\
Frequency Range & 4.5--6.5 GHz \& 8--10 GHz \\
Total observing time & 14 hours \\
Arrays & H168\\
Amplitude Calibrator & PKS B1934$-$638 \\
Phase Calibrator & PKS~B0823$-$500 \\
Frequency Resolution & 1 MHz \\
\hline\hline
\end{tabular}
\end{table}
\end{center}

\subsection{Observation details}

The Bullet cluster was observed for a total of 14 hours in the H168 array of the Australia Telescope Compact Array (ATCA) on July 30 \& 31, 2010, at 5.5 and 9 GHz. ATCA continuum mode with 1 MHz resolution \citep{2011MNRAS.416..832W} with all four Stokes parameters and 2048 channels was used for these observations. PKS~B1934$-$638 was used as the primary/amplitude calibrator and PKS~B0823$-$500 as the secondary/phase calibrator. Observation details has been tabulated in Table~\ref{Obs-sum}.

\subsection{Data analysis}

Data were analyzed using Multichannel Image Reconstruction, Image Analysis and Display (MIRIAD, developed by ATNF \citep{2011ascl.soft06007S,1995ASPC...77..433S}). Radio frequency interference induced bad data was excised. The secondary/phase calibrator was observed once every 60 minutes to keep track of variations in phase due to atmospheric effects. A description of our data reduction and calibration method is detailed in \citet{2016Ap&SS.361..255M}.

\begin{figure}
	\includegraphics[width=9cm]{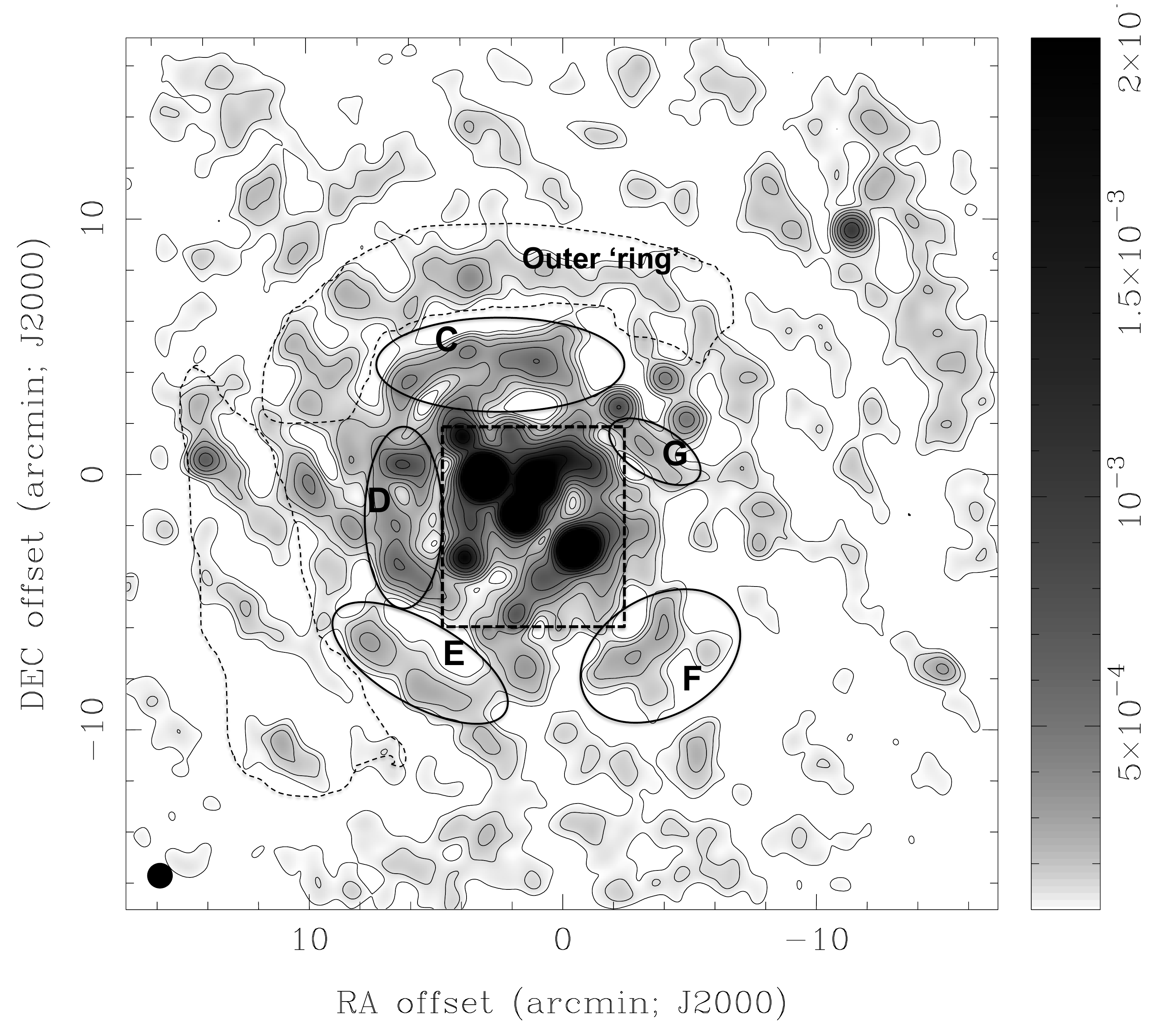}
\caption{5.5 GHz radio image of the Bullet cluster, observed with H168 array of the ATCA, with a synthesized beam 57$\arcsec\times$57$\arcsec$ (indicated as a circle in the bottom left corner). Specific regions are marked as A$-$G and flux densities are given in Table ~\ref{diffsrc}.}
    \label{cbandimage1}
\end{figure}

\begin{figure}
	\includegraphics[width=9cm]{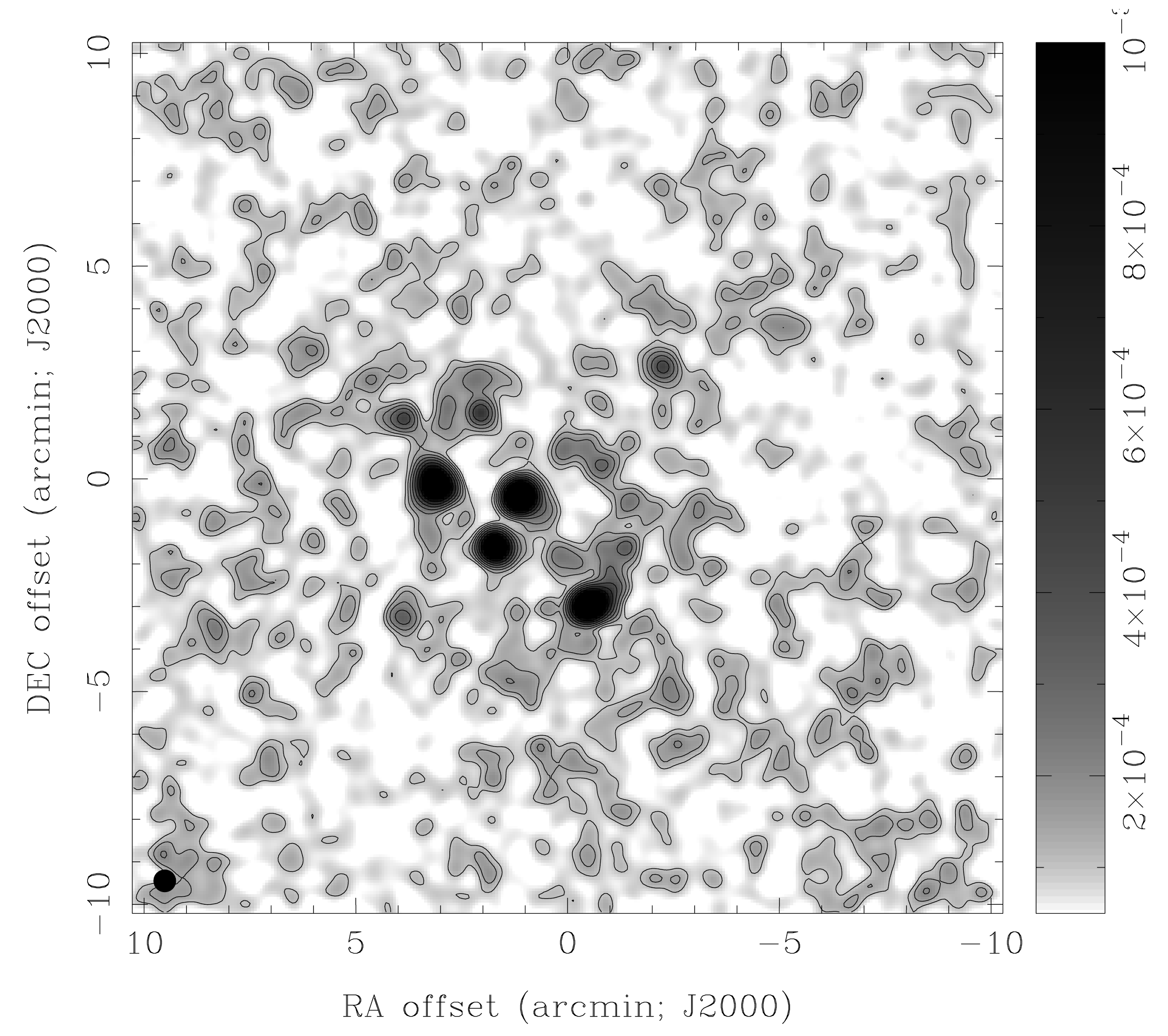} \caption{9 GHz radio image of the Bullet cluster, observed with H168 array of the ATCA, with a synthesized beam 30$\arcsec\times$30$\arcsec$ (indicated as a circle in the bottom left corner).}
    \label{xbandimage1}
\end{figure}

The images shown in Figure~\ref{cbandimage1} and \ref{xbandimage1} resulted from the  observations presented above. These images were made with antennas 1--5 of the ATCA in the H168 array. Figure~\ref{cbandimage1} was made using natural weighting. The extent of this image is larger than the figures in \citet{2014MNRAS.440.2901S}. Noise rms ($\sigma_{\mathrm{RMS}}$) is 20$\mu$Jy/beam.  The 5.5 GHz image has been convolved to 57$\arcsec\times$57$\arcsec$ with a PA of 0$\deg$. The intrinsic resolution in the first two 5.5 GHz images is the same as in \citet{2016Ap&SS.361..255M}. Regions A \& B, and five point sources are marked in the figure. Properties of the regions C, D, E, F \& G are given in Table ~\ref{diffsrc}. The two dashed areas indicate roughly the extents of the regions A and B. The dotted line represents the boundary between what we refer to as the peripheral regions or the periphery, and the `inner' part of the Bullet cluster. Figure~\ref{xbandimage1} was made with uniform weight, and the extent of this image is similar to the figures in \citet{2014MNRAS.440.2901S}. Noise rms ($\sigma_{\mathrm{RMS}}$) is 20$\mu$Jy/beam.

\subsection{Results}
It is evident from Figure~\ref{cbandimage1}, as well as from Figure 1 and 2 of \citet{2016Ap&SS.361..255M}, that there is significant diffuse emission in the periphery of the Bullet cluster, as observed at 5.5 GHz. We point out that the square dotted box in Figure~\ref{cbandimage1} is the extent of the 2.1 GHz image shown in \citet{2014MNRAS.440.2901S}, and, following them, \citet{2016Ap&SS.361..255M} examined diffuse and point sources within the dashed box. We concentrate here on diffuse emission outside this box only. 

The peripheral diffuse emission has been marked in Figure~\ref{cbandimage1}. In what follows, we use the coordinates RA: $06^{\rm h}58^{\rm m}30.0^{\rm s}$, DEC: $-55\deg56\arcmin30\arcsec$ as the reference, or the `center': this is the mid--point between the two X-ray peaks of the cluster.  Labeling of diffuse emission is done following \citet{2014MNRAS.440.2901S} and \citet{2016Ap&SS.361..255M}; since they label the relics `A' and `B', we label the first diffuse emission region `C' and so on. Because of this labelling, and to avoid confusion, we refer to the relic regions as Relic `A' and Relic `B', henceforth. 

Outside the dotted box, the diffuse emission has the appearance of `rings' surrounding the central region. These `rings' seem to have three layers, as marked with dotted arcs in Figure~\ref{cbandimage1}. The northern portions of the diffuse emission have arc-like structures, whereas those in the SE regions have elongated, straight features. Subdividing these regions is difficult, as many of them have `bridges', i.e. the diffuse emission features are connected through small regions of relatively low brightness. However, there is a clear gap between the first and second `layers' of peripheral diffuse emission. We have subdivided only the first `layer' of diffuse emission, and listed their properties in four sub-bands in the 4.5-6.5 GHz range, and two sub-bands in the 8-10 GHz range of frequencies.

\begin{figure}
	\includegraphics[width=8.5cm]{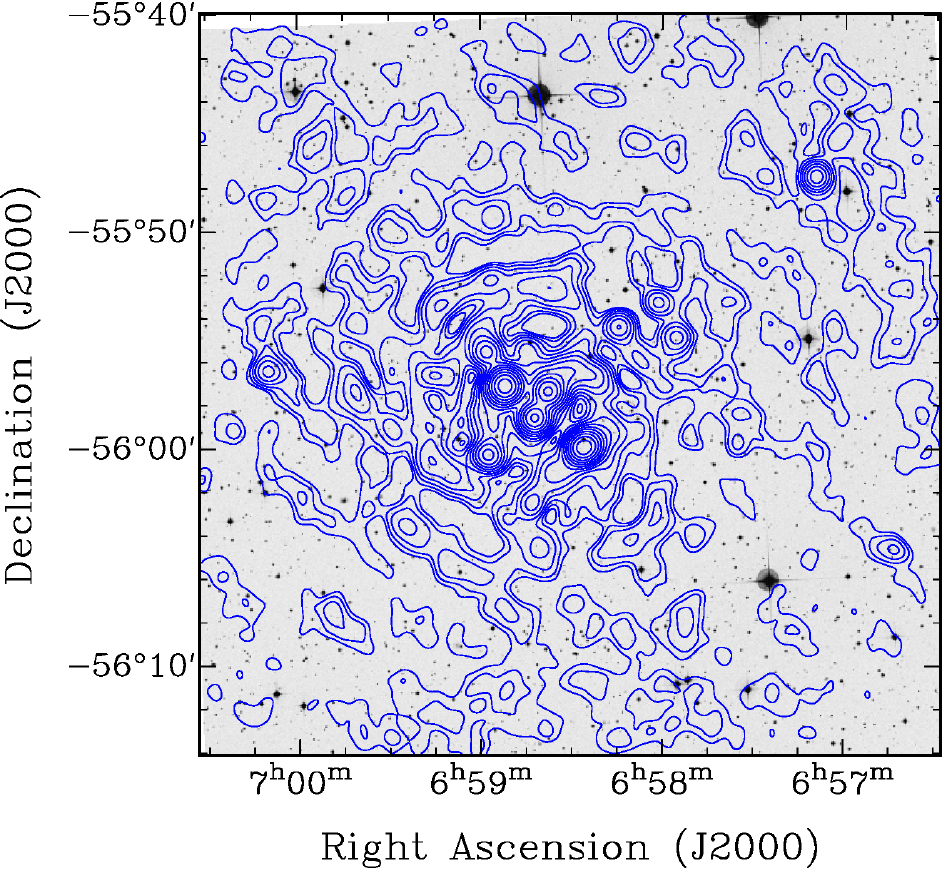} 
\caption{ Contours of the same image as in Figure~\ref{cbandimage1}, overlaid on the ESO DSS-1 optical image of the Bullet cluster, from the Oschin Schmidt Telescope on Palomar Mountain and the UK Schmidt Telescope. Contour levels start at 5$\sigma$ and increase by factors of $\sqrt{2}$, with the noise rms $\sigma$ = 20 $\mu$Jy/beam.}
    \label{overlaidimage1}
\end{figure}

Diffuse emission in the closer `ring' of emission in the N region is marked  as `C', that in the E region as `D', that in the SE region as `E', that in the SW region as `F', and that in the NW region as `G'. These regions have been clearly marked in Figure~\ref{cbandimage1}. It is evident from Figure~\ref{overlaidimage1} that these `rings' of diffuse emission do not follow the distribution of galaxies, and are morphologically different. Figure~\ref{overlaidimage1} clearly shows no correlation between diffuse emission, which is overlaid as blue contours, on the optical ESO DSS-1 image. 

\begin{table*}
\begin{threeparttable}
\begin{tabular}{lcclccccc}
\hline\hline
Source & \multicolumn{2}{c}{RA~~~~~~~~~~~~DEC} & $S_{\rm 2100~MHz}$ & \multicolumn{4}{c}{Int. Flux Density (mJy)} & \multicolumn{1}{c}{Int. Flux Density(mJy)} \\
Label & \multicolumn{2}{c}{(J2000)} &   & \multicolumn{4}{c}{in 5500 MHz band} & \multicolumn{1}{c}{in 9000 MHz band} \\ \hline 
& h:m:s & $^\circ$:$\arcmin$:$\arcsec$ & & $S_{\rm 4769~MHz}$& $S_{\rm 5223~MHz}$ & $S_{5751~MHz}$ & $S_{\rm 6193~MHz}$& $S_{\rm 9000~MHz}$ \\ 
\hline
Region C & 06:58:42.3 & $-$55:58:37.5 & 7.89 $\pm$ 0.10 & 4.66 $\pm$ 0.10 & 3.80 $\pm$ 0.15 & 3.23 $\pm$ 0.15 & 2.88 $\pm$ 0.20 & 1.20 $\pm$ 0.15 \\
Region D & 06:58:37.6 & $-$55:57:24.0 & 5.92 $\pm$ 0.10 & 2.17 $\pm$ 0.10 & 1.50 $\pm$ 0.10 & 1.15 $\pm$ 0.15 & 1.00 $\pm$ 0.10 & 0.19 $\pm$ 0.05 \\
Region F & 06:58:58.1 & $-$55:55:35.1 & 6.27 $\pm$ 0.04 & 2.14 $\pm$ 0.10 & 1.65 $\pm$ 0.15 & 1.30 $\pm$ 0.15 & $-$                      & 0.48 $\pm$ 0.05 \\
Region G & 06:58:57.8 & $-$56:00:20.1 & 1.14 $\pm$ 0.10 & 0.93 $\pm$ 0.05 & 0.81 $\pm$ 0.15 & 0.54 $\pm$ 0.20 & 0.47 $\pm$ 0.13 & 0.27 $\pm$ 0.08 \\
\hline\hline
\end{tabular}
\caption{Flux density of each of the demarkated regions given in mJy/beam.}\label{diffsrc}

\end{threeparttable}
\end{table*}

Observed multi-layered ring-like radio relic formation is very unusual and reported for the first time in this work. As we can see that the bullet event is currently on-going in the cluster, the shocks arising from this event definitely cannot reach a very large distance of few Mpc from the centre (outer radio rings Fig.~\ref{cbandimage1}). So, how this has formed? To understand this unique event, we need a time evolution study. Therefore, numerical simulations become inevitable.

\section{Understanding observed radio emissions from a simulated bullet cluster}\label{model}

\subsection{Simulation details}

Bullet like merger is one of the most complex events in the structure formation history \citep{Mastropietro_2008MNRAS}, thus cannot be understood well from only observations or from the ideal and controlled simulations. To unveil mysteries in such a rare and extreme object, we also need 3D realistic simulations in full cosmological environment. For this purpose, we have used {\sc ENZO}, an N-Body plus hydrodynamic AMR code \citep{Bryan_2014ApJS,O'Shea_2004astro.ph}. A flat $\Lambda$CDM background cosmology with the parameters of the $\Lambda$CDM model, derived from WMAP  \citep{2009ApJS..180..330K} has been used. The simulations have been initialized at redshift $z = 60$ using the \citet{1999ApJ...511....5E} transfer function, and evolved up to $z = 0$. An ideal equation of state was used for the gas, with $\gamma = 5/3$. Since, the emergence and propagation of shocks are the most important events in this study, we have thus captured the shocks very efficiently and resolved the grids adaptively where ever shock is generated and passes by using the method described in \citep{Vazza_2009MNRAS,Vazza_2011MNRAS}. In order to capture the correct energy distribution of Intra Cluster Medium, radiative cooling \citep{Sarazin_1987ApJ} 
and a star formation feedback scheme have been applied \citep{Cen1992ApJL}. A detail description of the simulation can be found in \citet{Paul_2017MNRAS,Paul_2018arXiva,John_2018arXiv}. 

Cosmological simulations were performed to create a sky realization of 128$^3$ Mpc$^3$ volume with 0.3 millions of particles and 64$^3$ grids at the root grid level. We have further introduced 2 child grids and inside the innermost child grid of $(32 Mpc)^3$ volume, we have implemented 4 levels of AMR based on both over-densities and the shock strength. With this set-up, we have achieved about 30 kpc spatial resolution and mass resolution of about 10$^{8} M_{\odot}$ at the highest resolved grids. For further details of the simulations and resolution studies of our numerical schemes, readers are suggested to go through \citep{Paul_2018arXiva,Paul_2017MNRAS,John_2018arXiv}.

\subsection{Simulated Bullet}
\label{simbullet1}

\begin{figure*}
\includegraphics[width=9cm]{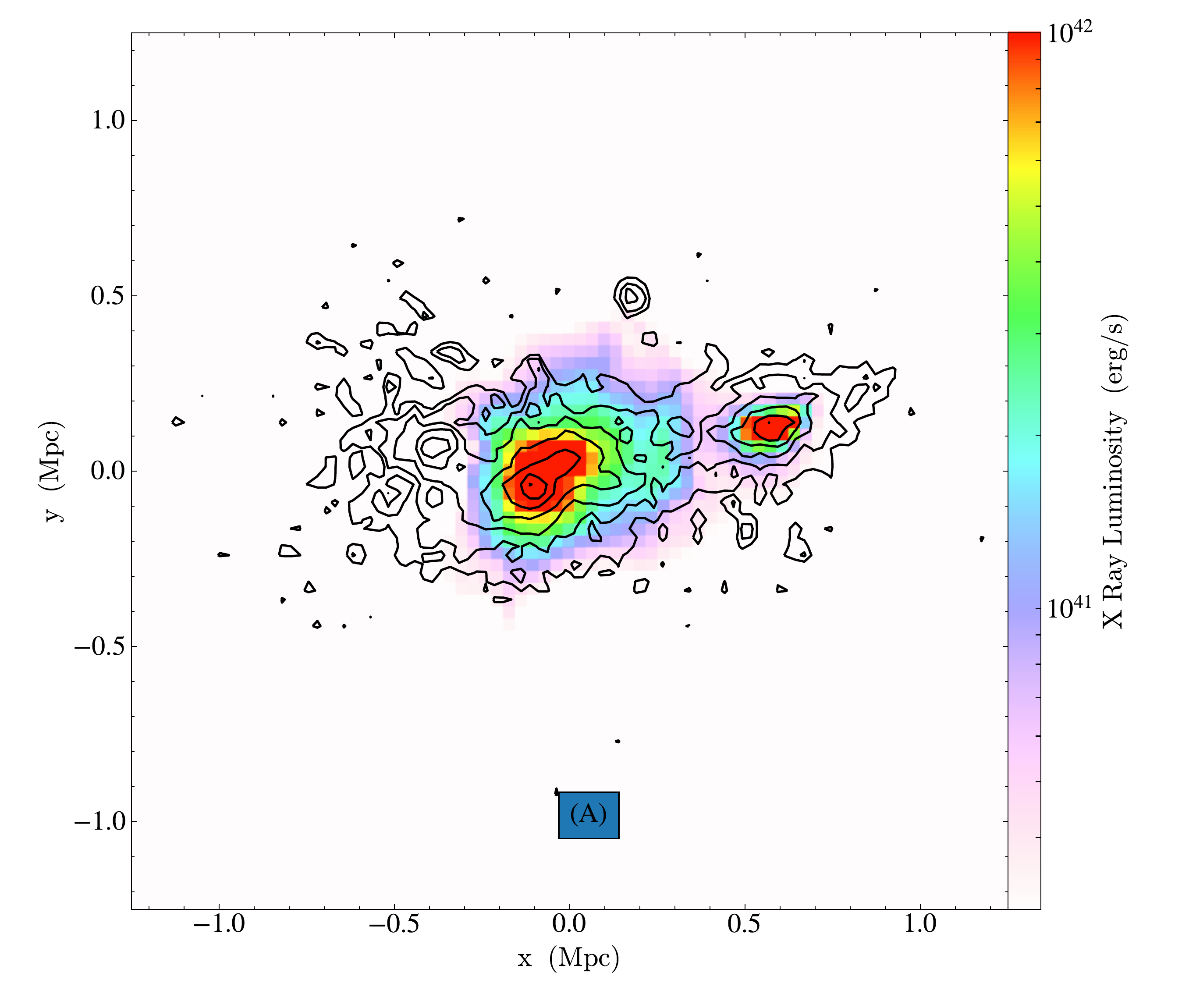}\hspace{-0.5cm}
\includegraphics[width=9cm]{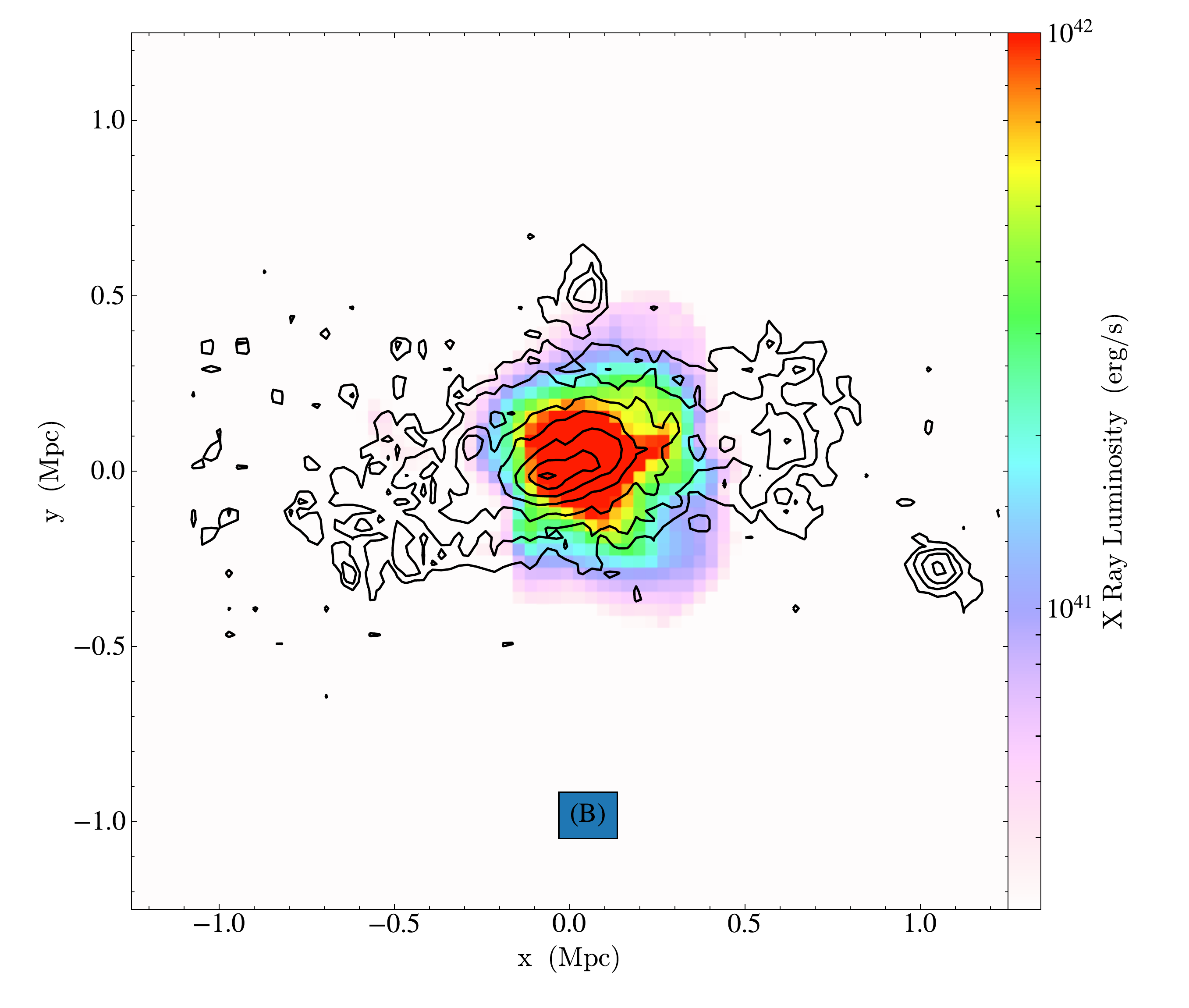} \hspace{-0.5cm}\\
\includegraphics[width=9cm]{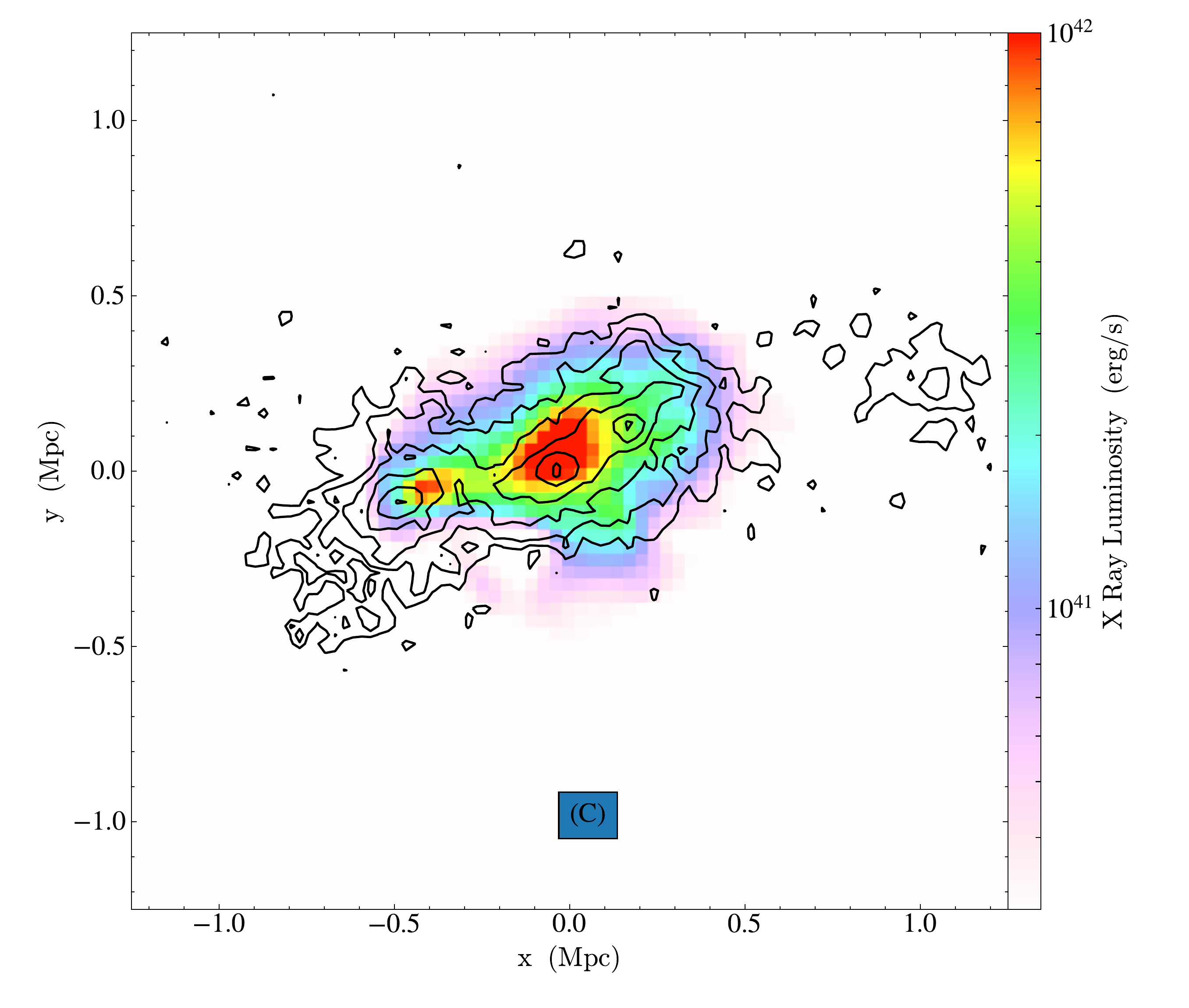}\hspace{-0.5cm}
\includegraphics[width=9cm]{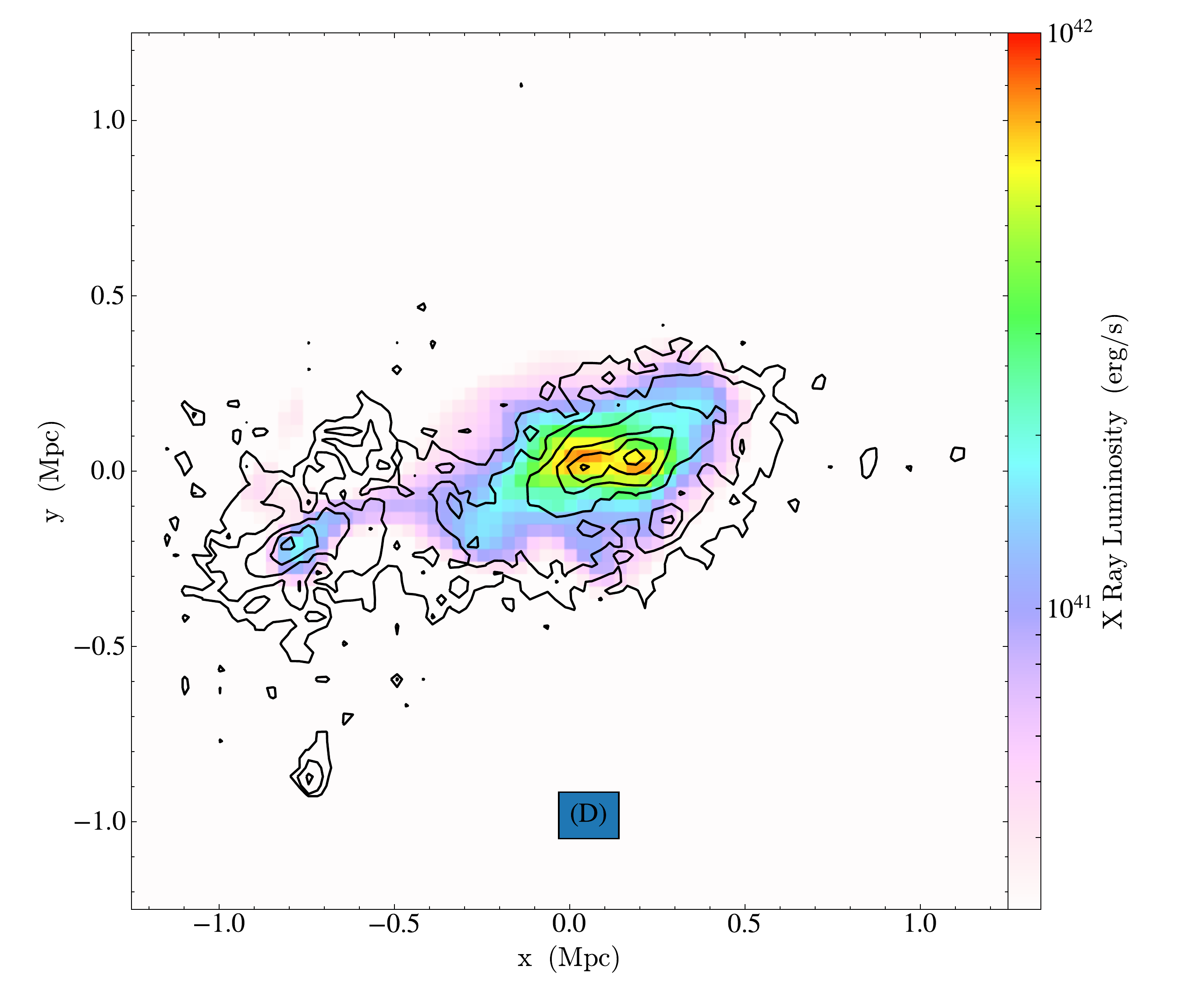}\hspace{-0.5cm}\\
\caption{A cluster merger event: Colour map shows the X-ray luminosity and black contours are the Dark matter. Panel 1-4 are representing pre-merger (redshift z= 0.315, merger = 0.295, bullet or core crossing at z=0.275 and end of post-merger state at z= 2.55 respectively.}\label{xry_dm}
\end{figure*}

Galaxy clusters are formed through hierarchical structure formation. From the merger tree produced for our simulation, we have found the biggest cluster with a mass more than 10$^{15}M_{\odot}$, and there are several clusters of mass $\sim$ 10$^{14}M_{\odot}$ and numerous $\sim$ 10$^{13}M_{\odot}$ objects. During the formation of the biggest cluster, it has gone through several mergers. These mergers are usually in the mass ratios of 1:10 to even 1:1. Equal ratio mergers usually yield extremely high energy, but in high ratio mergers, smaller groups are severely attracted by the bigger mass clumps and attain a velocity of high-supersonic (Mach~$M>3$) level similar to what has been observed in bullet cluster by \citet{Markevitch_2002ApJ}. In such cases, smaller groups ram into the larger groups and pass the cluster core extremely fast without getting merged \citep{Mastropietro_2008MNRAS} and it can be compared with the bullet hitting a target.

We have identified a merging event at about redshift $z=0.3$ in our simulation using the merger tree, where, a smaller group hits a big cluster. The  event is shown in Figure~\ref{xry_dm} using four panels showing the step by step process. Here, the small group of total mass of about $2\times 10^{14}M_{\odot}$ (left corner of Fig.~\ref{xry_dm}(A)) started falling into a much bigger cluster with total mass little more than $8\times 10^{14}M_{\odot}$ with a relative approaching velocity of 1800 km s$^{-1}$ as shown in Figure~\ref{xry_dm}(A). The system has begun interacting at the redshift z=0.315 and nearest approach is at z=0.295 (Fig.~\ref{xry_dm}(B)). After the first passage, the smaller group rams past like a bullet with a velocity about 2700 km s$^{-1}$ creating a bow shock in the front of the bullet (Fig.~\ref{hist} and Fig.~\ref{xry_dm}(C)). The shock Mach number is around $M=3.5$, indicating a shock velocity of about 5000 km s$^{-1}$. In any normally merging clusters, it is impossible to see even a shock of Mach number $M=2$ inside the hot cluster core. We have also studied the velocity structure of the simulated cluster and plotted as a histogram in Figure~\ref{hist}. We found two distinctly separated peaks with the main cluster having a median speed of about 1155 km s$^{-1}$ and the bullet is having a median of about 2670 km s$^{-1}$. The bullet travelled a Mpc within 0.42 Gyr i.e. with an average speed of about 2400 km s$^{-1}$.

Further, to confirm the other reported parameters of the observed bullet cluster, we studied the DM and baryon distribution in our simulated bullet. As reported, possibly bullet is the first example of a cluster where baryons get separated out from dark matter \citep{Markevitch_2002ApJ}. Figure~\ref{xry_dm}(a)$-$(d), show the evolution of baryons and dark matter in our simulated cluster. It shows that initially the peak of dark matter (see the contours) and the X-ray emitting gas were coinciding (Fig.~\ref{xry_dm}(A)). But, after the smaller clump passes the bigger one, the non-interacting dark matter went far ahead, while due to thermal energy loss and viscous drag, X-ray emitting baryons lagged behind, just like the observed bullet cluster (Fig.~\ref{xry_dm}(C)). Baryonic matter gets severely attracted back by the DM core ahead. Finally, baryons catch the dark matter core once again after almost a Gyr (see (Fig.~\ref{xry_dm}(D)). This is the first report of clear separation of DM and baryon in a full set up simulation with $\Lambda$CDM cosmology. Such event has been observed in (128 Mpc)$^3$ simulation, indicating a possibility of finding more such events if larger volume is simulated or if an extensive sky search is done.

\begin{figure}
\includegraphics[width=9cm]{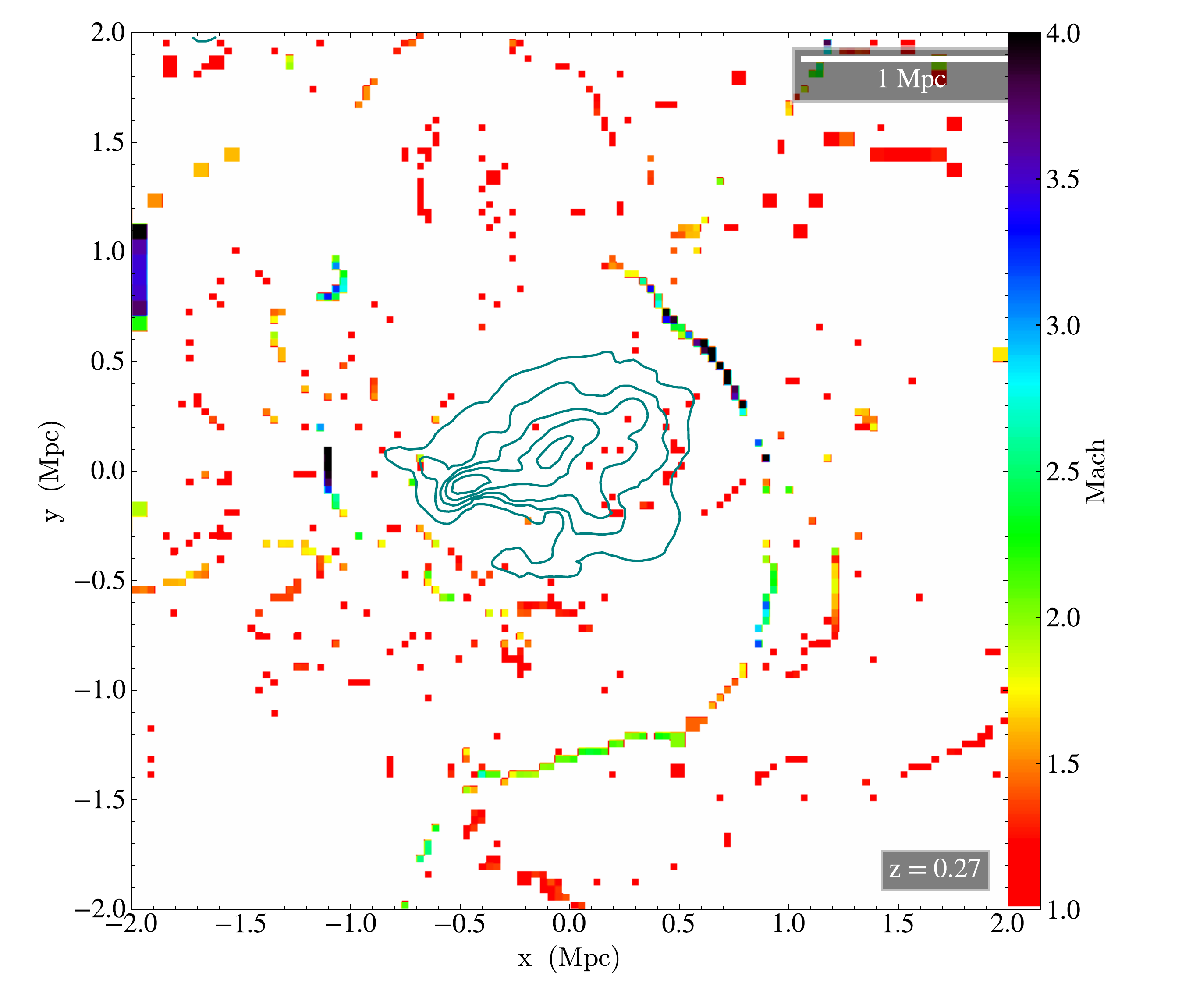}
\caption{Mach number of the shocks are plotted in colour with black density contours for an area of (4 Mpc)$^2$.}\label{shock-mach}
\end{figure}

\begin{figure}
\includegraphics[width=8.5cm]{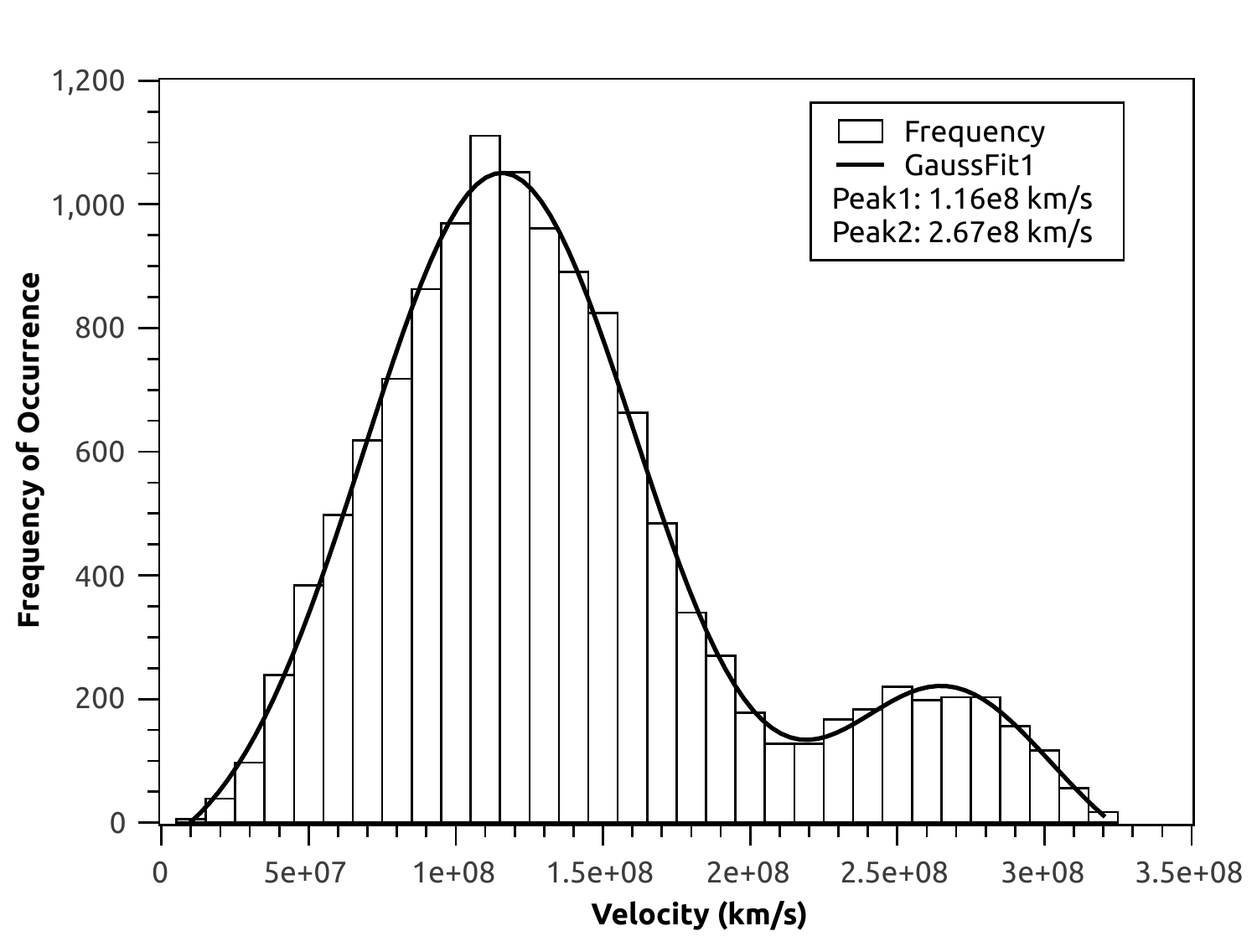}
\caption{Histogram of the velocity distribution in the simulated bullet cluster.}\label{hist}
\end{figure}

\subsection{Shocks and turbulence in the modelled `bullet' cluster}\label{shock-turb}

\begin{figure}
\includegraphics[width=9cm]{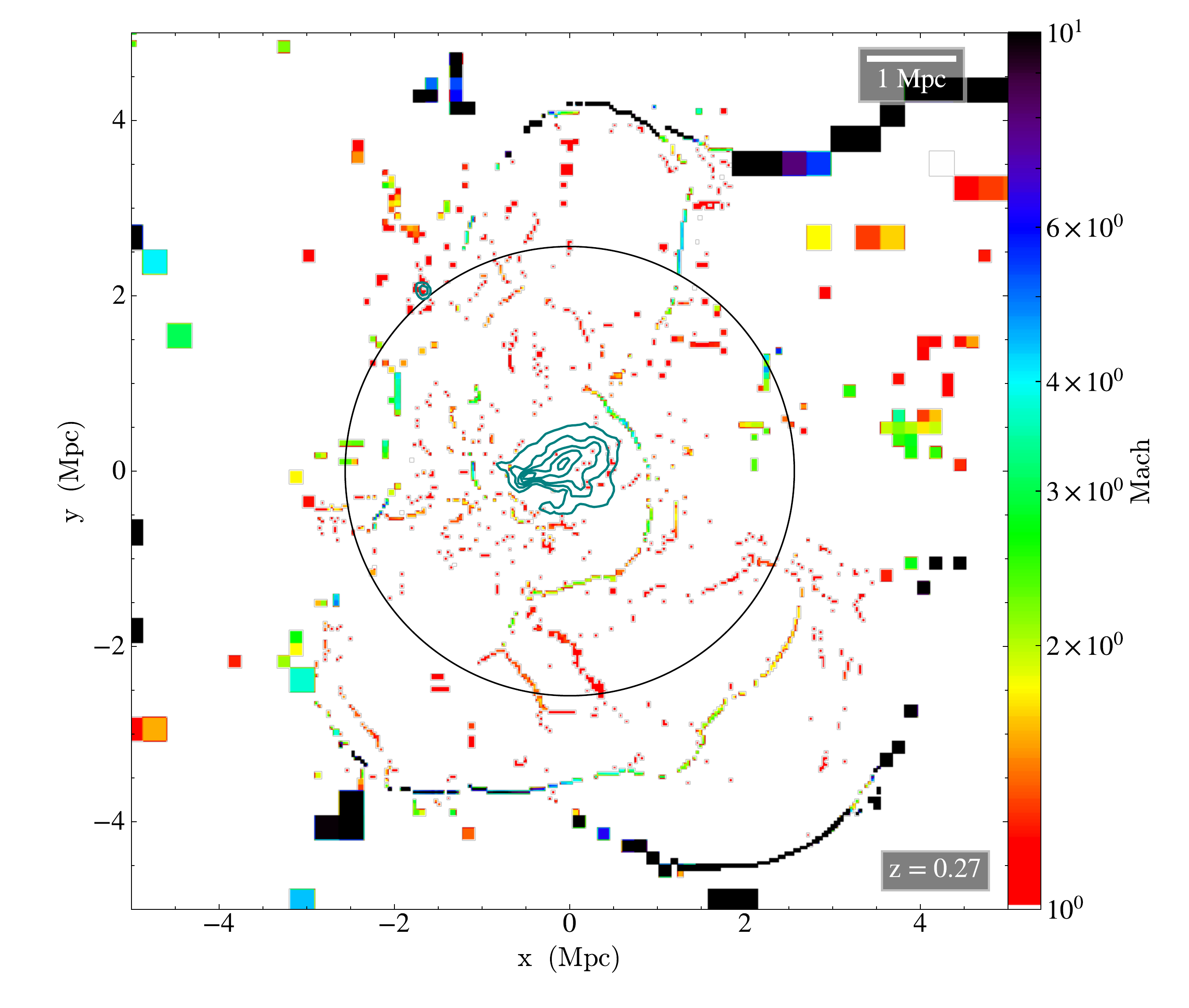}
\includegraphics[width=9cm]{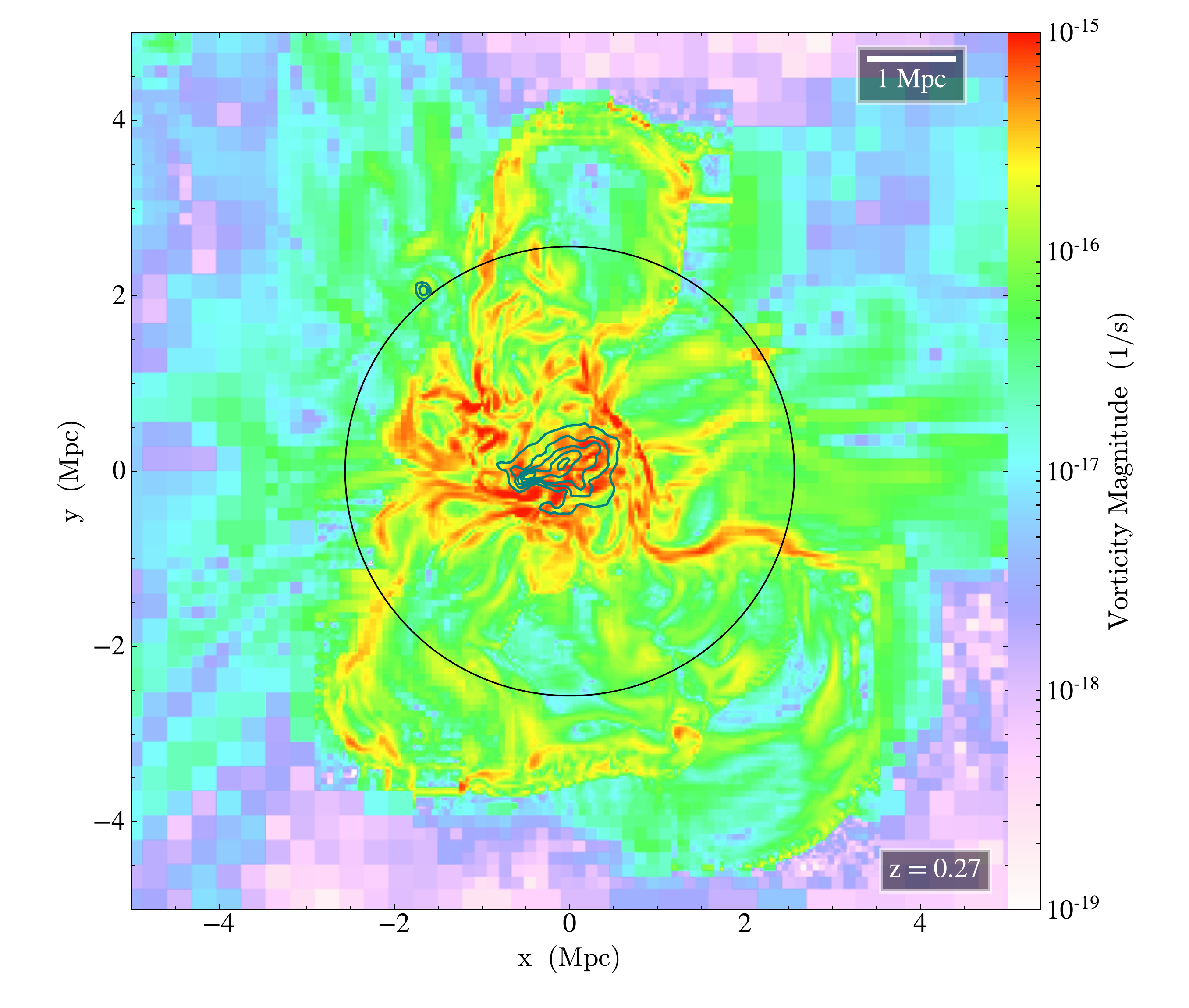}
\caption{{\bf Panel 1:} Mach number of the shocks are plotted in colour with black density contours for an area of (10 Mpc)$^2$. {\bf Panel 2:}  Vorticity magnitude ($\lvert{\omega}\lvert$) of plotted for the same area.} \label{Shock-vort-fig}
\end{figure}

We have produced a video for the time evolution of thermal shocks for a period of few Gyrs (see the supporting supplementary material). It shows the evolution of the thermal properties much before and after the bullet event. While, the bullet has produced a strong shock ($M=3.5$) in the medium (see Fig.~\ref{shock-mach}), shocks from earlier mergers (before the bullet event has taken place) produced multiple layers of shocks that are seen to move with a supersonic speed (see the video of temperature shock), creating spherical shock waves. These shocks are the efficient in situ particle accelerator (through DSA) and can become visible in radio waves. Since, these merging shocks last for more than few Gyrs \citep{Paul_2011ApJ}, with an average speed of 1000 km/s they are observed to reach to the periphery of the clusters and meet the accretion shock. Further, a snapshot of the simulation plotted for the Mach number at redshift z=0.27 (see Fig.~\ref{Shock-vort-fig}, Panel 1), clearly shows the multiple supersonic shocks around the cluster and the bow shock that was seen ahead of the bullet in Figure~\ref{shock-mach} is placed at a very short distance from the cluster centre. Mach number of the shocks outside the virial radius (shown as the black circle), is beyond 5 and reached more than 10 for the merger shocks those meet the accretion shocks, indicating a possible show up in radio.

We have also plotted the vorticity ($\vec{\omega}=\vec{\nabla} \times \vec{v}$) in the cluster after the bullet has passed the centre (see Fig.~\ref{Shock-vort-fig}, Panel 2). Vorticity being a proxy to the solenoidal turbulence in the medium \citep{Vazza_2017MNRAS,Miniati_2015Natur}, it is very important for the cluster radio halo emissions through turbulent re-acceleration. We have found that the bullet has induced a high vorticity of 10$^{-15}$ s$^{-1}$ which corresponds to a turbulent velocity of more than 2000 km s$^{-1}$,  in the vicinity of its movement. Rest of the cluster is at a lower level of turbulence with vorticity magnitude about an order lower. We can also observe that the vorticity is also higher in the region behind the shocks that emerged out of some previous merger indicating a better efficiency of turbulent re-acceleration and magnetic field amplification through turbulent dynamo.

\subsection{Radio emission computation}
\label{radioemcompute}

Radio emission takes place in a magnetised medium if relativistic charged particles, usually the electrons are available. The ICM is known to host magnetic field of $\mu$G order \citep{2004IJMPD..13.1549G} that can be achieved through turbulent dynamo model applied on primordial seed magnetic field \citep{Subramanian_2006MNRAS}. But, it needs a high degree of turbulence in the ICM. A saturation of magnetisation can be obtained when medium becomes fully turbulent, usually Kolmogorov type $E(k) \propto k^{-5/3}$ (where $k$ is wave number). A simple equipartition condition can be considered between magnetic energy density $\frac{B^2}{8\pi}$ and the kinetic energy density $\rho \epsilon_{turb}$ in this condition \citep{1998MNRAS.294..718S}. A detailed description of the method can be found in \citet{Paul_2018arXiva}. In Figure~\ref{Shock-vort-fig}, Panel 2, we saw that the `bullet' event has generated a considerable amount of turbulence in the medium. Implementing above model, we have found a few times of $\mu$G magnetic field in our simulated `bullet' system.

Further, to obtain the radio emissions due to the bullet event, we need an abundance of relativistic electrons that will gyrate in the said magnetic field to emit through synchrotron process \citep{Longair_2011}. The known best particle acceleration mechanism active in large scale structures that can elevate thermal electrons to relativistic energies are the diffusive shock acceleration  (DSA) \citep{Drury_1983RPPh} and the turbulent re-acceleration \citep{Brunetti_2001MNRAS,2011MNRAS.412..817B}. The synchrotron power spectrum in this medium will be determined by the electron energy spectrum from either DSA or the TRA. As we have noticed that both the shocks and the turbulence are available at a very high level in our simulated bullet (see section~\ref{shock-turb}), they could easily supply the required high energy particles to produce radio emission in the system.

Thermally distributed ICM particles after shocks acceleration (DSA) get converted to a power-law energy distribution $N(E) dE \propto E^{-\delta} dE$ \citep{1983RPPh...46..973D}. Where, $\delta$ is the spectral index of electron energy with value 2 and more steeper. With turbulent re-acceleration (TRA), the electron energy power-law takes up the form 
\begin{equation}
\left(\frac{dn_e}{dE_e} \right) = \frac{3P_A\,c}{4 S(E_{\rm max})^{1/2}}\,E_e^{-\delta}
\end{equation}

where, $P_A$ is the part of the total turbulent power going into the Alfven waves and E$_{max}$ is the maximum available electron energy. A simple assumption would be to consider a fully developed Kolmogorov type turbulence that makes the $\delta$ to be $\frac{5}{2}$ \citep{2016JCAP...10..004F}. Here, we did not include synchrotron ageing model, which may lead to a slightly different spectrum of radio emission. 

Synchrotron Radio emission is then calculated using the standard synchrotron emission formula 
\begin{eqnarray}
 \frac{d^{2}P(\nu_{obs})}{dVd\nu} &=& \frac{\sqrt{3}e^3B}{8m_e c^2}\,\int_{E_{\rm min} }^{E_{\rm max}}dE_e\,F\left(\frac{\nu_{obs}}{\nu_c}\right)\,\left(\frac{dn_e}{dE_e} \right)_{\rm inj}
\end{eqnarray}
where $F(x)=x\int_x^\infty K_{5/3}(x')dx'$ is the Synchrotron function, $K_{5/3}$ is the modified Bessel function, and $\nu_c$ is the critical frequency of synchrotron emission, 
\begin{equation}
\nu_c =  {3}\gamma^2{eB}/{(4\pi m_e)} = 1.6\, ( {B}/{1 \mu \rm G})({E_e}/{10 \rm GeV})^2 ~ \rm GHz 
\end{equation}

$\frac{dn_e}{dE_e}$ is the injected electron energy spectrum determined by either be DSA \citep{Hong_2015ApJ} or the TRA mechanism \citep{2016JCAP...10..004F},. For a detail account of the computation of radio emissions for our study, please see \citet{Paul_2018arXiva}.

\subsection{Simulated radio map from the modelled `bullet'}\label{sim-res}

Implementing above models for computing magnetic field and electron energy spectrum, radio emission has been computed on snapshots taken from our hydrodynamic simulations. For computing radio emission from DSA electrons, shocked cells are identified. Computation is done on the grid parameters of these cells and a proper weight has been used to nullify the effect of complicated resolution pattern of an AMR simulation. The computing formulae has been implemented on each grids and different output parameters have been visualized as slice or projection plots of few Mpc scales as required. 

\begin{figure*}
\includegraphics[width=8.8cm]{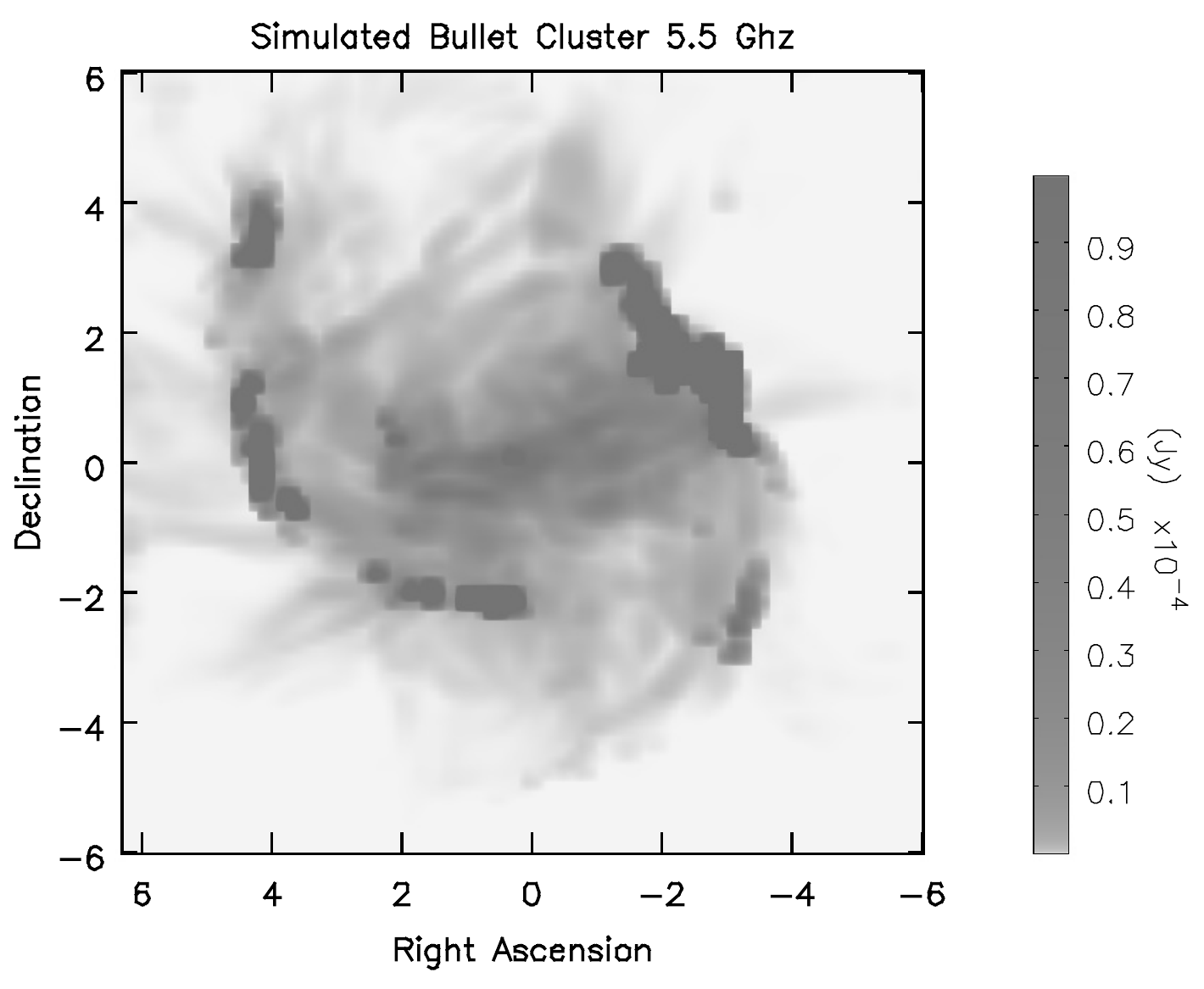}\hspace{-0.3cm}
\includegraphics[width=8.8cm]{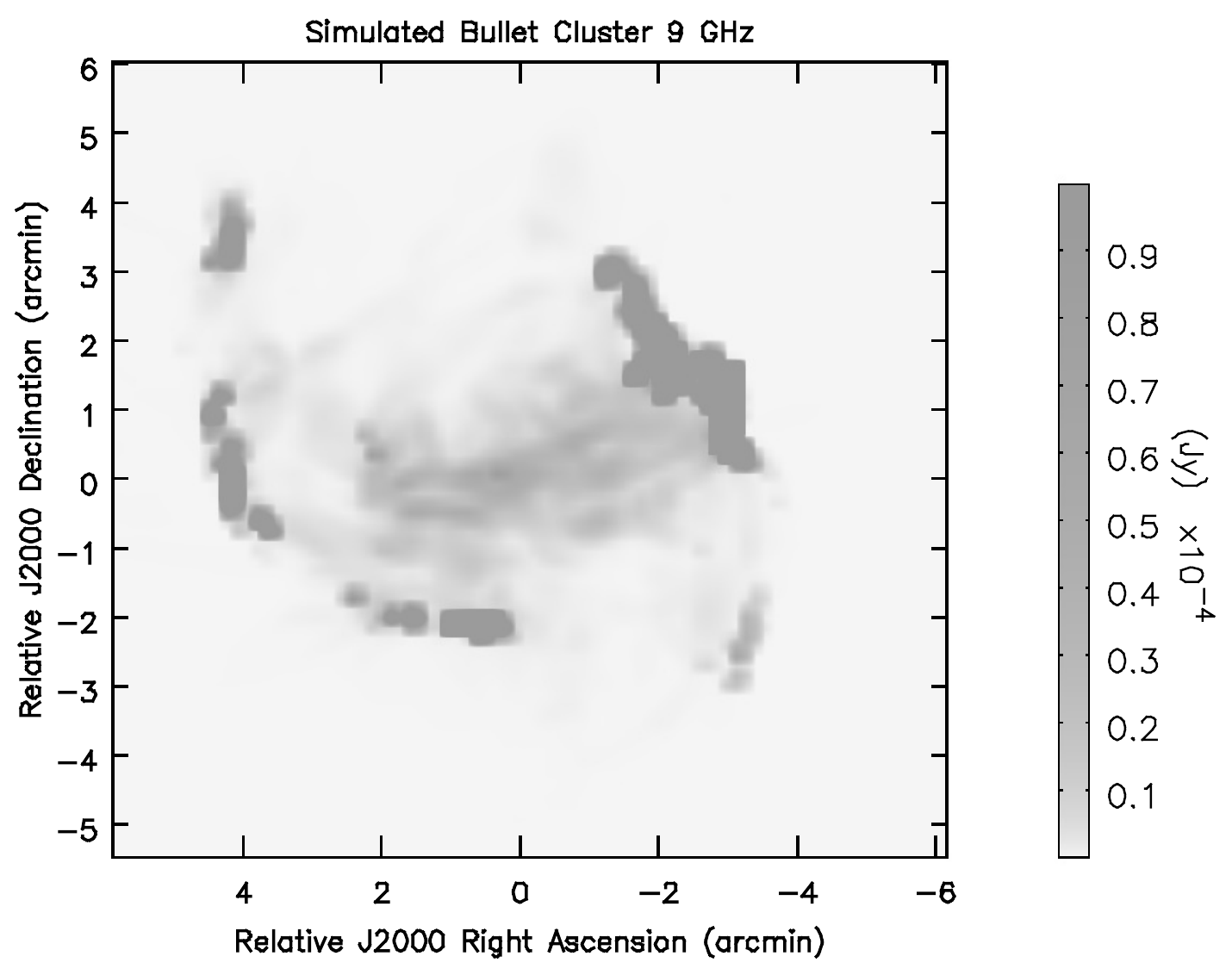}\hspace{-0.5cm}\\
\includegraphics[width=8.8cm]{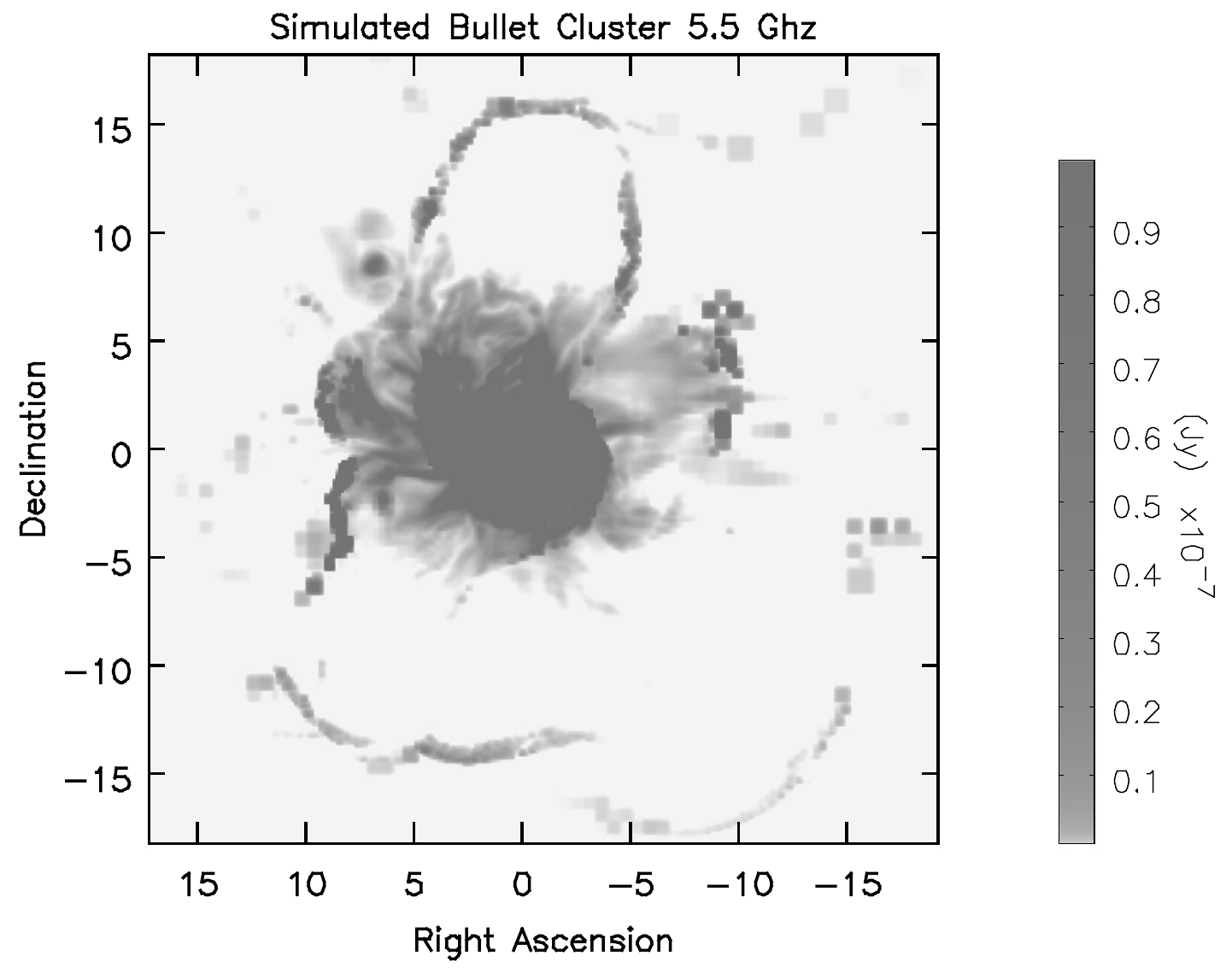}
\hspace{-0.3cm} \includegraphics[width=8.8cm]{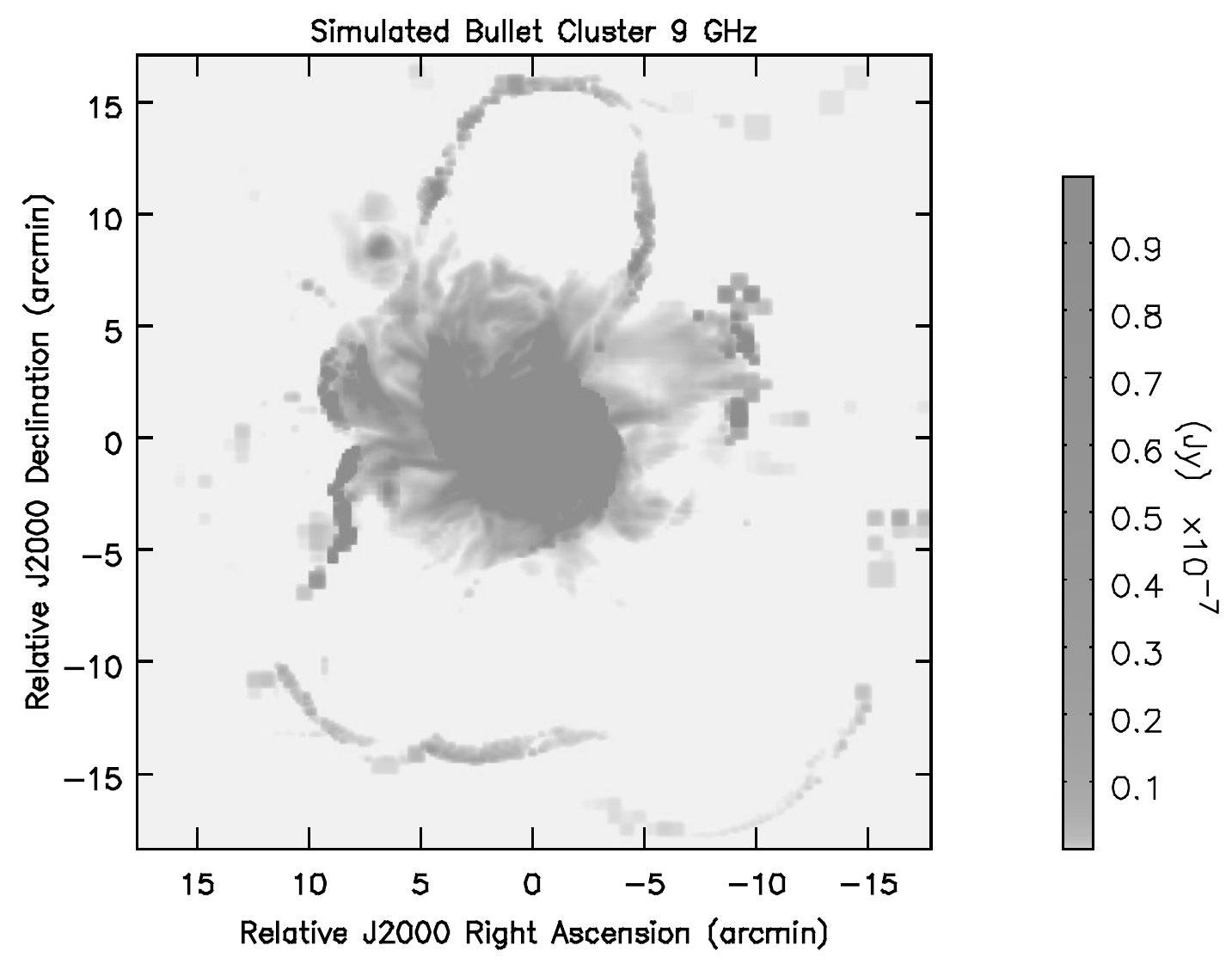}
\hspace{-0.5cm}\\
\caption{{\bf Upper row} Slice plots of central $12\arcmin \times 12\arcmin$ area showing DSA + TRA radio flux on earth in Jy at redshift z=0.275 i.e. at bullet cluster like phase of merger. Radio maps are made at 5.5 GHz (Panel 1) and 9 GHz (Panel 2). {\bf Lower row} Same for a $32\arcmin \times 32\arcmin$ area.}\label{rad-map}
\end{figure*}

Radio emission from this merging system clearly shows that the bullet is associated with a radio shock front (see Fig.~\ref{shock-mach}). The shock fronts are mostly having flux of 10s of $\mu$Jy in 5.5 GHz and almost half an order below in magnitude at 9 GHz. When we have plotted the same cluster with (10 Mpc)$^2$ area in Figure~\ref{Shock-vort-fig}, Panel 1 and Figure~\ref{rad-map} Panel 3--4, it shows more outer shocks in radio waves. These shocks are far away from the cluster and are perpendicular to the merging axis. The order of flux is very low i.e. about few 10s nano Jy. With a beam convolution of $57\arcsec \times 57\arcsec$ at 5.5 GHz and $30\arcsec\times 30\arcsec$ at 9 GHz, we observe total flux density to be varying in the range of sub mJy to a few mJy for different regions as observed with ATCA (Table~\ref{diffsrc}). Morphologically, they produce multiple `ring' like shock structures similar to the Figure~\ref{cbandimage1} at 5.5 GHz. As seen in the Figure~\ref{Shock-vort-fig}, Panel 1, these are associated with merger shocks from earlier mergers. From the distance of these shocks from the centre of the cluster a simple estimation can be done about the look-back time for the actual merging event. Further, these mergers had taken place along another axis than the bullet, more than a few Gyr back. From the video and the radio image (in Fig.~\ref{rad-map}, Panel 3--4) it is quite evident that the outer most relic has originated from the interaction of merger and accretion shocks that has Mach number $M>10$ and internal 2--3 layers of shocks are the multiple merger shocks with Mach number $M\eqsim2$ (see Fig.~\ref{Shock-vort-fig}, Panel 1).

\section{Discussion and conclusion}
\label{last}

We have presented the discovery of  peripheral diffuse radio emissions surrounding the Bullet cluster, in multiple ring-like structures at 5.5 and 9 GHz, using the ATCA. There are several (marked as `outer ring') regions of these peripheral emissions that are greater than two virial radii away from the centre of the merging system. Origins of observed peripheral diffuse emission in a system like the bullet cluster, as reported in this paper, is not very straightforward to deduce, especially when these peripheral emissions are further away from the cluster virial radius. To explain such a unique and complicated morphology, here, for the first time, we present a simulation of a bullet like event in a full cosmological environment with N-Body plus Hydrodynamics as well as thermal physics. The structures observed in the bullet cluster have been nicely reproduced in our simulations (see Section~\ref{radioemcompute}). Various simulated `bullet' parameters such as the bullet velocity, shock strength, DM-baryon distribution as well as the radio flux and extent of peripheral diffuse emission is strikingly similar to the observed values. 

\subsection{Comparing bullet event}

Our model shows, a bullet is a violently merging slingshot that ramps pass the main cluster core with an average velocity of 2700 km s$^{-1}$ creating a bow shock of Mach number $M=3.5$ which corresponds to a shock velocity of about 5000 km s$^{-1}$. We have also found a clear separation of baryonic matter of bullet from the DM, similar to as reported in \citep{Markevitch_2002ApJ}. So, we report that the bullet event is a viable process in $\Lambda$CDM cosmology i.e. a very high velocity of 2700 km s$^{-1}$ and the temporal separation of DM and baryons both are the possible scenarios. Since, it has been observed in merely a (128 Mpc)$^3$ simulated volume, we expect that this may not be an extremely rare event or posses any challenge to $\Lambda$CDM cosmology as of now as doubted by \citet{Lee_2010ApJ,Kraljic_2015JCAP}. Both the authors have concluded from their DM only simulations. \citet{Lee_2010ApJ} has also mentioned that a velocity of above 1800 km s$^{-1}$ is impossible in $\Lambda$CDM, but our velocity histogram in Figure~\ref{hist} clearly shows a bullet with average velocity of 2700 km s$^{-1}$. It also shows that the bullet has travelled a Mpc within a time period of 0.42 Gyr (redshift $z=0.315-z=0.270$) with an average velocity of 2400 km s$^{-1}$. So, we attribute this discrepancy of findings to their DM only simulations which completely lacks the information about the transient behaviour of baryons inside a DM environment.

\subsection{Understanding `rings' of peripheral emission}

This bullet event creates the first layer of radio ring due to emergence of high Mach bow shocks as observed in Figure~\ref{rad-map}, Panel 1 and 2 in 5.5 GHz and 9 Ghz, the same feature as it is observed in the Bullet cluster. Observed relics are close enough that one of them -- Relic `A' -- even has a `radio bridge' connecting it to the radio halo. 

The observed diffuse emission at about 2 Mpc away from the cluster merger centre (see Fig.~\ref{cbandimage1}, Relic C, D, F and G) possibly be a radio relic from an earlier merger shock as shown in the simulated image in Panel 3, Figure~\ref{rad-map}. \citet{2015MNRAS.449.1486S} have pointed out the evidence for a second shock, but Relic `B' has formed possibly at the same time as the bullet as, the shocks emerged from the oscillation of bigger cluster that was disturbed by the bullet itself. This is the reason that it is still significantly closer to the cluster centre. 

Shock waves due to large-scale structure accretion on to galaxy clusters have possibly been observed by \citet{2006Sci...314..791B} in the cluster/merger Abell 3376, where they also report ring-like structures surrounding the cluster, which are $\sim$ 1--2 Mpc away from the cluster centre. A two-layer shock with a possible combination of merger and accretion has also been reported by \citep{Bagchi_2011ApJ} in PLCK G287.0+32.9.
We report another `ring' of emissions (Fig.~\ref{cbandimage1}, outer 'ring' ) at the outermost region of our map. There is no report of a third shock in the Bullet cluster preceding to our work. This farthest one is almost two virial radius away, at the accretion zone. \citet{2001ApJ...563..660F} show that diffuse radio emission from accretion shocks can be a factor of a few order less than merger shocks. Mach numbers of accretion shocks can be significantly greater than those of merger shocks, since the gas in cluster outskirts has not been heated by shocks before\citep{2000ApJ...542..608M,2006MNRAS.367..113P,2008MNRAS.391.1511H}. At the same time, the surface brightness of any peripheral diffuse emission would fade away with time (see, e.g. Fig. 2 in \citet{2011JApA...32..509H}). The observed outer ring is also very fainter, indicating a possibility that these `rings' of diffuse emission are caused by the accretion shocks in the cluster outskirts. But, the density of available high energy charged particles become extremely low in the peripheral region, so for a detectable amount of emission, it will need another power engine. So, a third model can be invoked as the interaction of merger shocks with the accretion shock that can re-energise the medium near to the accretion shock.

A more clearer picture can be observed from our simulated bullet like cluster in Section~\ref{radioemcompute}. We found 2-3 layers of shocks in our simulations. The farthest one is almost two virial radius away and the intermediates are almost at the virial radius. Looking at the time evolution in the supplementary video, it can be clearly seen that the merger shock from a previous merger meets the peripheral accretion shock at the same instance when bullet structure has been formed. This interaction seen to increase the temperature substantially and can significantly re-energise the ambient medium as well as shock compresses the magnetic field and produce radio emissions as described briefly in Section~\ref{intro} and observed in our simulations (see section~\ref{sim-res}). 

\section*{Acknowledgements}
The Australia Telescope Compact Array is part of the Australia Telescope which is funded by the Commonwealth of Australia for operation as a National Facility managed by CSIRO. This research has made use of the services of the ESO Science Archive Facility, specifically, the ESO Online DSS Archive. SP would like to thank DST INSPIRE Faculty Scheme (IFA-12/PH-44) for supporting this research. SM is grateful to Mark Wieringa, Robin Wark, Maxim Voronkov and the research staff at ATCA/ATNF for guidance and help with understanding the CABB system. We thank Mark Wieringa for useful discussions. Observations and analysis were made possible by a generous grant for Astronomy by IIT Indore. SC acknowledges support by the South African Research Chairs Initiative of the Department of Science and Technology and National Research Foundation of South Africa (Grant no. 77948). SP, RSJ and PG are thankful to the Inter-University Centre for Astronomy and Astrophysics (IUCAA) for providing the HPC facility. Computations described in this work were performed using the publicly-available \texttt{Enzo} code (http://enzo-project.org) and data analysis is done with the yt-tools (http://yt-project.org/)




\bibliographystyle{mnras}
\bibliography{bullet9} 

\label{lastpage}
\end{document}